# Phonon Kinetics at the Solid to Liquid Phase Transition

Alex Davie[1], Farah Vandrevala[2], Sara Dampf[3], Yanting Deng[1], Deepu K. George[1], Eric D. Sylvester[4], Timothy Korter[3], Erik Einarsson[2,5], Jason B. Benedict[4], Andrea G. Markelz[1]

[1]*Department of Physics, University at Buffalo, Buffalo, New York, United States*
[2]*Department of Electrical Engineering, University at Buffalo, Buffalo, New York, United States*
[3]*Department of Chemistry, Syracuse University, Syracuse, New York, United States*
[4]*Department of Chemistry, University at Buffalo, Buffalo, New York, United States*
[5]*Department of Materials Design and Innovation, University at Buffalo, Buffalo, New York, United States*

**Abstract**
Terahertz time domain spectroscopy (THz TDS) is used to measure the melting kinetics of fructose molecular crystals. Combining single crystal anisotropy measurements with density functional calculations we assign the phonon frequencies and interrogate how specific phonons behave with melting. While nearly all the low frequency phonons continuously red shift with heating and melting, the lowest energy phonon polarized along the *c*-axis blue shifts at the melting temperature, suggesting an initial structural change immediately before melting. We find that the kinetics follow a 3D growth model with large activation energies consistent with previous differential scanning calorimetry (DSC) measurements. The large activation energies indicate multiple H-bonds must break collectively for the transition. The results suggest the generality of the kinetics for molecular crystals and that THz TDS with picosecond resolution could be used to measure ultra-fast kinetics.

## I. Introduction

Phase transitions of molecular crystals are important across a range of technologies. For example, crystallization in pharmaceuticals can lead to reduced drug efficacy. In the field of energetic materials, the melt transition determines the energy release rate. Calorimetry is the dominant method for characterizing phase transitions, from which one can readily extract the enthalpy of the transition. However, monitoring the kinetics is challenging due to the time resolution limits of fast differential scanning calorimetry (FDSC).

A number of studies have demonstrated that crystal phonons can be used to monitor structural transitions. Previously, optical techniques have been used for melting measurements of covalently bonded crystalline materials, but there are no reports of phonon anharmonicity and melting kinetics reported for molecular crystals where weak hydrogen bonding is responsible for the crystal structure, leading to phonons in the far infrared or THz range.

Terahertz time domain spectroscopy (THz TDS) has been used as a standard technique for characterizing molecular crystal phonons. THz TDS spectral acquisition can be on the nanosecond time scale and thus can provide an alternative to FDSC to extract phase transition kinetics. However, the majority of THz TDS phonon studies of phase transitions have focused on the amorphous to crystalline transition (1-7). This is mainly due to the challenges of performing a melting measurement. These challenges include a spectroscopic configuration to ensure temperature control and spectral integrity as the sample loses structure. We demonstrate the use of THz TDS to monitor structure and kinetics of melting using single crystal fructose.

Fructose, along with several other molecular crystals, can superheat beyond the so-called melting temperature, allowing for kinetic measurements using DSC. Those measurements indicate that a high number of hydrogen bonds are involved in the melting of fructose and that there may be structural change at the transition (8).

Here we use terahertz spectroscopy on single crystal fructose to examine how the crystal phonons evolve during the phase transition. We compare the single crystal anisotropic absorbance to DFT calculations to identify the phonons. We find phonons polarized along the *a* and *b* axes monotonically decrease in frequency as the temperature approaches melting and this trend continues with melting. Contrary to this, for the lowest frequency phonon, with symmetry along the *c* axis, the red shifting halts at the melting temperature and the phonon then blue shifts and diminishes with amplitude with the loss of structure, suggesting an atomic reconfiguration during the transition before the loss of structure.

We analyze the melting using an Avrami-Erofeev model where the integrated absorption line of a b-polarized phonon is used to determine the fraction of crystallinity. We find that the model is well fit by a nucleus growth model with n = 3, consistent with the DSC results. Further we find a melting rate $k = .0230\ min^{-1} \pm .0024\ min^{-1}$ which gives kinetic parameters entirely consistent with DSC measurements. The results support that multiple H-bonds must collectively break for the phase change to occur. The consistency with DSC results, suggests the methodology can be used for general kinetics studies of phase transitions.

## II. MATERIALS AND METHODS

*Crystal Growth and Polishing*

Single fructose crystals were grown using the seeded growth method. 40 grams of pure fructose powder was dissolved in 5 ml of water and 48 ml of 90% ethanol at 60ºC. Fructose seeds with lateral dimensions ~ 2 mm were grown by leaving the saturated solution at 40ºC, checking regularly for growth. Seeds were then strung and suspended in clean saturated solution at 40ºC for 1-2 days, checking for growth at regular intervals resulting in single crystals ~ 1 cm$^3$, with easily identifiable facets. Crystals were then hand polished along a specific crystal face with a surface area > 5 mm x 5 mm and with less than 500 μm thickness to ensure the maximum absorbance was within the dynamic range of the THz TDS system. Polished crystals were then indexed using X-ray diffraction. Our single crystals of fructose were found to be orthorhombic with a space group of $P2_12_12_1$ and lattice constants $a = 8.088$ Å, $b = 9.029$ Å, $c = 10.034$ Å. Three crystals were indexed to identify both the polished face and the facets.

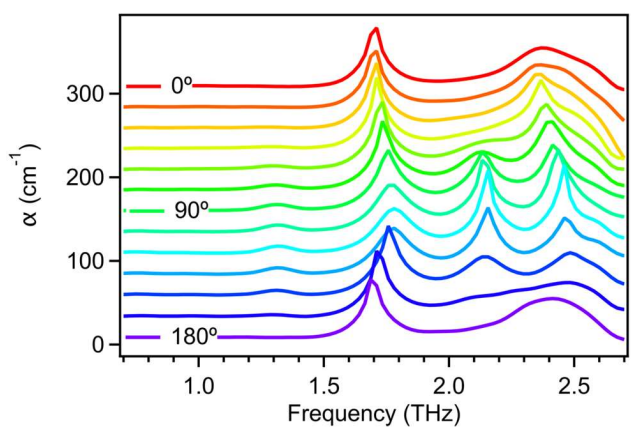

**Fig. 1** Anisotropic absorption coefficient for (101) faced single crystal fructose. The angle of the THz polarization relative to the [100] axis is shown on the curves. The resonant frequencies are determined by Gaussian fits and listed in Table 1. The amplitudes follow a $cos^2\theta$, with the polarization angle $\theta$ relative to the [100] axis.

*Anisotropic THz Spectroscopy*

The anisotropy of all samples was first characterized using THz TDS. Polished crystals were mounted on apertures and placed in a sample rotator and oriented relative to the THz polarization. Transmission was referenced to transmission through an empty aperture. The absorbance spectra were analyzed by applying curve fits to each orientation. Combining the crystal indexing and the polarization dependence we determine the orientation of the dipole transition for the resonances. Two crystals gave the full assignment. One with a (101) face with the (010) edge identified provided initial assignment of vibrations along the *b*-axis (see Figure 1). The additional resonances could not be

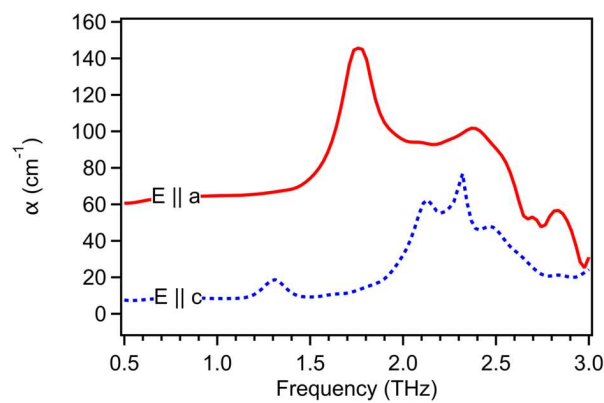

**Fig. 2.** Anisotropic absorption coefficient for (010) face single crystal fructose. Phonons polarized along the b-axis do not appear in this data set and the *a*-axis and *c*- axis polarized phonons can be identified.

Table 1

| DFT Calculated Frequencies | | Calculated Intensity | Measured Frequencies | Polarization |
|---|---|---|---|---|
| cm$^{-1}$ | THz | km/mol | THz | |
| 53.2053 | 1.5951 | 0.01 | 1.30 +/- .03 | c |
| 64.6351 | 1.9377 | 4.21 | 1.67 +/- .03 | b |
| 70.4335 | 2.1115 | 3.28 | 1.76 +/- .03 | a |
| 76.041 | 2.2797 | 4.25 | 2.13 +/- .03 | c |
| 85.103 | 2.5513 | 3.42 | | b |
| 100.924 | 3.0256 | 23.93 | | c |

assigned as both *a*-axis and *c*-axis phonons will be present for the THz polarization perpendicular to the *b*-axis for this crystal. Absorbance spectra with crystals with (010) polished faces and (001) edge identified provided the full assignment of the resonances observed with the THz TDS system. When multiple anisotropic THz measurements on different samples are paired with each samples respective X-ray diffraction measurements, we are able to unambiguously assign these modes to a specific crystal direction. By doing this we are able to characterize the absorbance spectrum of fructose from 0.5 – 2.5 THz. Examining Fig. 1, the anisotropic absorption is plotted for different polarizations relative to the crystal axes for (101) face crystal. 0 deg corresponds to the field along the [010] axis where there is an obvious resonance at 1.7 THz. 90 deg from this the polarization is along the [101] direction, and resonances are seen at 1.3 THz, 1.76 THz, 2.13 THz and 2.45 THz. In Fig. 2 we show the polarization dependent absorption for a (010) face crystal with 45 deg corresponding to the field along the [001] direction and resonances are seen at 1.3 THz and 2.13 THz. 135 deg corresponds to the field along the

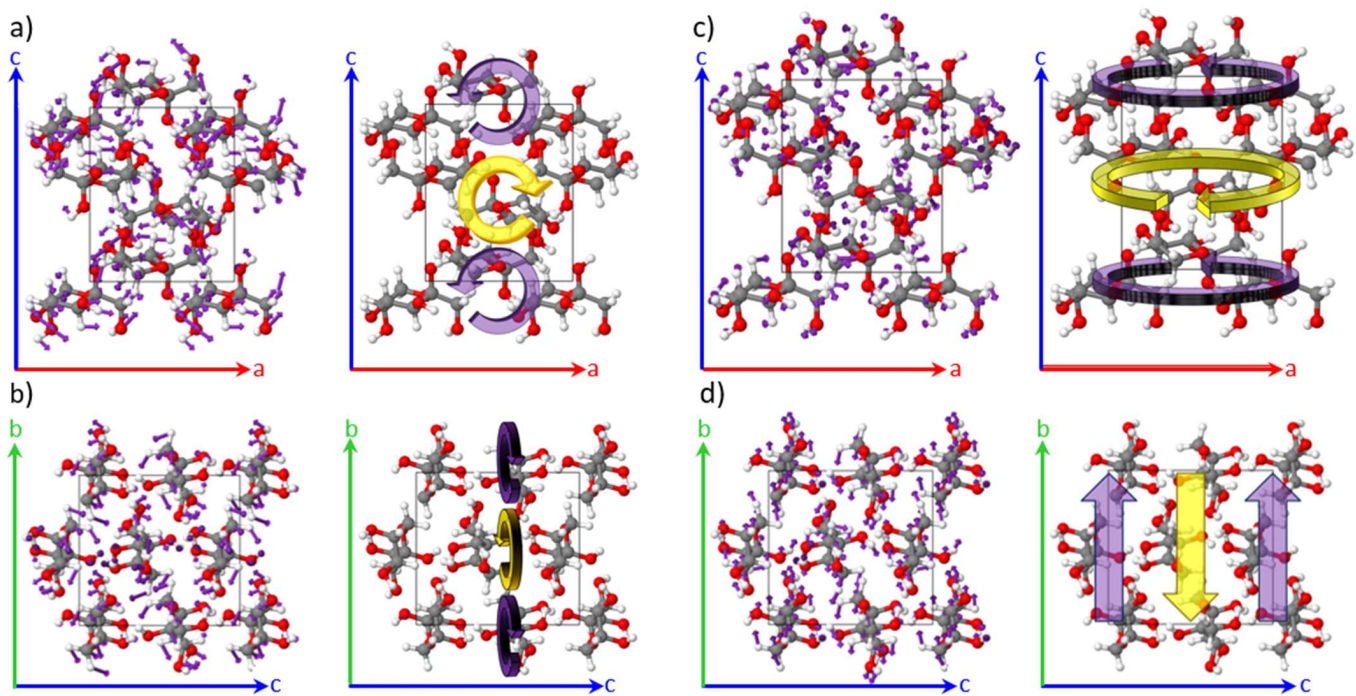

**Figure 3. Displacement vectors for the first four optically active phonons determined by DFT calculations.**

[100] direction and resonances are seen at 1.76 THz and 2.45 THz. An additional (101) indexed crystal was also measured to verify the results shown here. By combining the three measurements we can assign the 1.3 THz and 2.13 THz resonances along the *c*-axis, the 1.70 THz along the *b*-axis and the 1.76 and 2.45 THz along the *a*-axis. The frequency ordering and polarization dependence compared to the DFT allow us to assign the specific motions of the phonons for the resonances, shown in Fig. 3 and Table 1.

*DFT Calculations*

All DFT calculations were performed using the CRYSTAL17 software package utilizing periodic boundary conditions to properly simulate the environment of a crystalline solid (9). The calculations employed the Perdew-Bruke-Ernzerhof (PBE) density functional with the addition of Grimme's London dispersion correction (D3) with Becke-Johnson damping and inclusion of a three-body dispersion term (10-13). The atom-centered Gaussian-type function basis set, Ahlrichs' VTZ was used with added polarization functions (VTZP) (14, 15) and obtained from the Basis Set Exchange (16). To ensure numerical accuracy, a pruned DFT integration grid with 99 radial points and 1454 angular points was used with 125k points in the irreducible Brillouin zone, and the overlap-based truncation tolerances for the Coulomb and exchange integrals were set to $10^{-10}$, $10^{-10}$, $10^{-10}$, $10^{-10}$, $10^{-20}$. The crystal structure was allowed to optimize freely to an energetic minimum with a convergence of $\Delta E < 10^{-8}$ hartree and was restrained only by space group symmetry ($P2_12_12_1$).

Good agreement of the resulting structure with the previously published low-temperature structure was achieved (17). This optimized structure was then used in a vibrational frequency calculation with an energy convergence set to a stricter $\Delta E < 10^{-10}$ hartree for numerical differentiation of the potential energy surfaces associated with atomic displacements. The vibrational analysis of the crystalline system yielded harmonic vibrational frequencies and accompanying infrared absorption strengths calculated through the Berry phase approach. The infrared intensities also included the three-dimensional Born charge tensor polarization components for each normal mode of vibration (18, 19). Normal mode atomic eigenvector displacements were visualized with the Jmol application to understand the characteristic motions of each vibration (20).

*Temperature Dependent THz Spectroscopy*

Temperature dependent terahertz measurements were performed on Advantest's TAS7500TS commercial THz-TDS system. A custom sample heater paired with a temperature controller was used to heat the sample. The metal sample holder has two apertures and is attached to a heater. The sample was placed on a quartz substrate placed in one of the apertures between the heater and a resistance temperature detector sensor that was attached to the sample holder with thermal cement. The temperature was slowly raised from room temperature to the melting point of fructose, taking measurements at regular intervals. Once at the melting point, the temperature was held constant and repeated measurements were taken. There seemed to be inconsistencies related to the spectra from the Advantest system. The spectra seemed to be stretched or compressed in several cases. To correct this, a simple calibration was done with each measurement. This calibration consisted of applying curve fittings to the resonances in the Advantest data. Since the samples were characterized beforehand it is possible to plot the frequency of the resonances from the characterization versus the frequencies obtained from the Advantest system to get a linear relation. This linear relation was then used to correct the frequency domain of the Advantest data.

III. RESULTS AND DISCUSSION

*Anisotropic THz Spectroscopy*

Figure 1 shows the absorption coefficient for a (101) face fructose crystal. The terahertz 0º polarization is along the *b*-axis. At 0º there is a sharp resonance at 1.70 THz and a smaller amplitude broad resonance at 2.30 THz. As the sample is rotated about the [101] direction the feature at 1.70 THz is no longer visible and new features at 1.30 THz, 1.76 THz, and 2.13 THz can be seen. The data indicates that 1.70 THz phonon is polarized along the *b*-axis. We can also determine that the features at 1.30 THz, 1.76 THz, and 2.13 THz are from motions along either the *a*-axis or the *c*-axis, but using this sample alone it is difficult to assign them to either axis. In order to assign the 1.30 THz, 1.76 THz, and 2.13 THz resonances to either the *a* or *c*-axis, a second indexed crystal was used in which the terahertz electric field is incident on the (010) face. In this instance the *b*-axis is not in the plane of the electric field, and thus we will not see the modes along that axis. In figure 2 we show measurements for a (010) faced fructose crystal. The terahertz 0º polarization was initially along the *a*-axis and the feature at 1.76 THz is visible, whereas rotating the sample by 90º with the polarization along the *c*-axis at 1.30 THz and 2.13 THz. From the measurements on the [010] faced sample, we can conclude that the 1.30 THz and 2.13 THz modes correspond to motions along the *c*-axis and the 1.76 THz mode is due to motions along the *a*-axis. The resonance at 1.70 THz is also not present in the (010) sample, which is expected assuming the motions are along the *b*-axis. An additional (010) sample was used to confirm the results. We then turn to the DFT calculations to unambiguously assign the first four resonances in fructose in the terahertz regime to specific lattice vibrations.

Table 1 shows the results of the DFT calculations ordered by frequency (from lowest to highest) along with the polarization of the dipole transition. The experimental results with the same frequency ordering are shown. The frequency ordering shows agreement with the polarization of the vibrations, allowing us to assign the atomic displacements of the measured phonons. The DFT frequencies are somewhat blue shifted from the experimental results. Figure 3 shows the displacement vectors for the assigned lattice vibrations.

*Temperature Dependent THz Spectroscopy*

Temperature dependent THz measurements were done on multiple fructose single crystals in order to isolate different modes and study their evolution through the melting phase transition. A temperature run for the *b*-axis polarized 1.70 THz, is shown in Figure 4a and *c*-axis polarized 1.30 THz phonon in Fig. 4b. The central frequency, amplitudes and linewidths are extracted using Lorentzian curve fitting. The temperature dependence of the central frequency is shown in Fig. 5, (data points colored according to time). Fig. 5a shows that the 1.70 THz mode continuously decreases in frequency with increasing temperature with a net shift of 0.085 THz, until the sample finally lost structure. This red shifting as the temperature is expected, as the lattice force constants decrease with expansion of the unit cell volume until it melts and loses structure.

Fig. 5b shows the temperature dependent terahertz measurements for lowest frequency 1.30 THz mode. Initially the phonon red shifts by 0.039 THz, much like the 1.70 THz mode. However, as the crystal reaches the nominal melting temperature of 374K the phonon begins to blue shift. The change suggests a structural rearrangement initially before melting. Once the sample reached 374 K the mode began to blue shift with a total amplitude of 0.014 THz. The anomalous blue shift we observe is mostly unexplained and there is little literature that describes such a shift. In a paper by Liavitskaya et al.(8) calculations showed that fructose had a surprisingly high melting activation energy. They determined that since the energy to break a hydrogen bond is much lower than the activation energy, then there must be some reorientation of molecules during melting to account for the difference. The red and blue shift that is observed for the 1.30 THz mode might be due to the reorientation described in the Liavitskaya paper. We note that the *a*-axis polarized phonon at 1.76 THz and *c*-axis polarized phonon at 2.13 THz continuously red shift with melting. The lowest-energy phonons are most intermolecular. It is possible that the blue shift is only seen in the lowest energy phonon along the *c*-axis, which may indicate a long-range structural reorganization along the c-axis before melting.

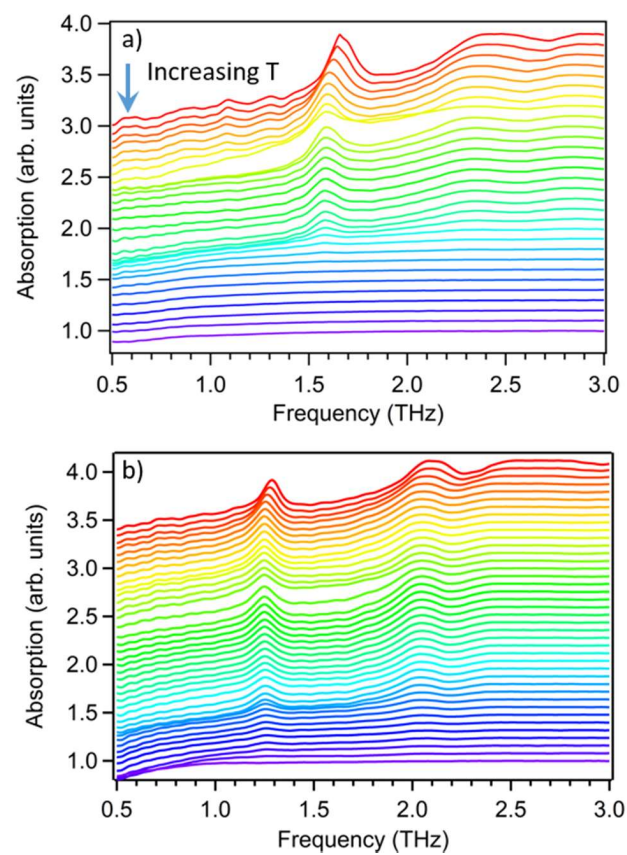

**Figure 4. Temperature dependence spectra of fructose single crystals. a) (101) face crystal oriented with THz polarization along [010]. B) (010) face crystal oriented with THz polarization along the [101] direction. The 1.3 THz phonon blue shifts at the melting temperature whereas the other peaks continuously red shift.**

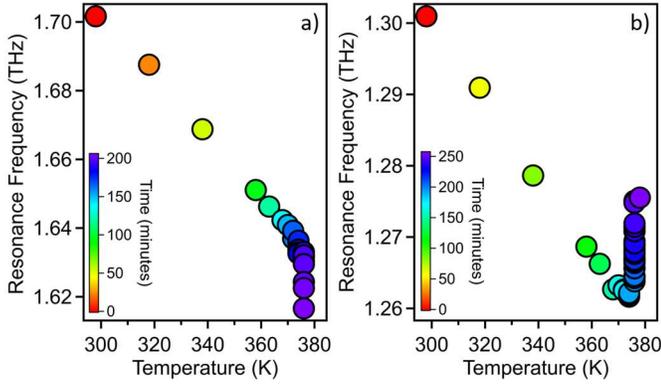

**Figure 5.** Resonant frequency as a function of temperature for the a) 1.7 THz and b)1.3 THz resonances. The 1.7 THz resonance continuously red shifts with increasing temperature and melting, whereas we see the 1.3 THz line blue shifts close to the melting temperature and as the temperature is held at 376 K.

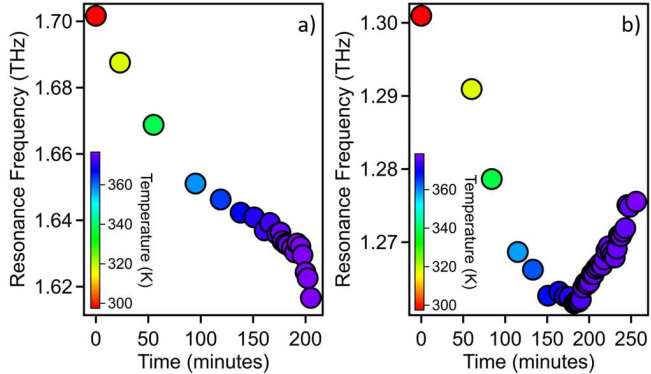

**Figure 6.** Time dependence of the resonant frequency through the melting transition. a) 1.70 THz resonance and b) 1.30 THz resonance. The onset of the blue shift for the 1.30 THz occurs at the melting temperature, whereas the 1.70 THz resonance continuously redshifts with temperature. The red shift appears to accelerate at melting. We use the 1.70 THz resonance to examine the melting kinetics.

To examine the kinetics and compare our results to the DSC studies, we plot the crystalline ratio as a function of time and fit to the Avrami-Erofeev model described by

$$r_c(t) = e^{-(k(t-t_o))^n} \quad (1)$$

where $r_c$ is the crystalline fraction, $k$ is the melting rate, $t_0$ is the onset time for melting, and $n$ is a constant dependent on geometry. To determine $r_c$ from the data we take an approach similar to Hedoux et al. where they use integrated Raman scattering resonances to track the fraction of crystallinity and showed excellent agreement with microcalorimetry results (21). To avoid the kinetics associated with the possible structural change, we use the 1.70 THz resonance as this peak continuously red shifts with increasing temperature and time.

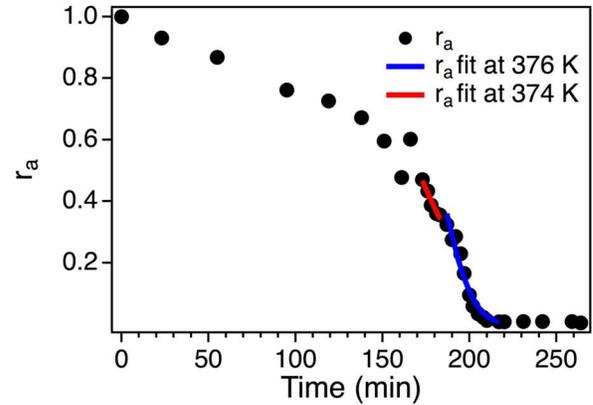

**Figure 7.** A plot of the amorphous ratio determined from the spectra of the 1.70 THz mode. The red line is a fit of the Avrami-Erofeev to the 374 K data points and the blue line is a fit to the 376 K data points.

For the spectral fitting used to attain the time and temperature dependence of the resonance shown in Fig. 5, there is as a smooth baseline associated with the amorphous phase absorption. This baseline is subtracted from each spectrum taken at time *t*, and the absorbance intensity between 1.4 and 1.9 THz is integrated, which is a sufficient region to account for both the line width and the red shift.

$$I_{1.7}(t) = \int_{1.4\,THz}^{1.9\,THz} (Abs(\nu,t) - Abs_{baseline}(\nu,t))d\nu \quad (2)$$

We then define the crystalline fraction as $r_c = I_{1.7}(t)/I_{1.7}(0)$ as the sample is entirely crystalline at $t = 0$. The resultant $r_c(t)$ is shown in Fig. 7. We then fit $r_c(t)$ for the region of time where the temperature is held constant to Eq. 1 for different values of *n*. We find $n =3$ gives the best fit, supporting the melting follows a random nucleation model. At 376 K we find that $k = .0230\ min^{-1} \pm .0024\ min^{-1}$ and $t_0 = 143.14\ min \pm 5.21\ min$ and at 374 K we find that $k = .010788\ min^{-1} \pm .00144\ min^{-1}$ and $t_0 = 88.202\ min \pm 12\ min$. To estimate the activation energy, we combine these results and use an Arrhenius temperature dependence of the melting rate:

$$k(T) = k_o e^{\frac{E_A}{RT}} \quad (3)$$

where $k_o$ is a prefactor, $E_A$ is the activation energy and $R$ is the gas constant. We can find an estimate for the activation energy of fructose of $441.8 \frac{kJ}{mol}$. This value is significantly larger than the reported value from Liavitskaya et al. but further supports their conclusion that fructose has an unexpectedly high activation energy. As the 374 K fitting is over a narrow range, leaving the estimate of the activation energy somewhat uncertain, we examine if our extracted melting rate is consistent with the parameters Liavitskaya et al. extracted from DSC measurements. Using their value for the activation energy of 220 kJ/mol at 105°C and our value of the melting rate, we find a prefactor $ln(k_o) = 62.4$, in near perfect agreement with the DSC determined value.

## DISCUSSION

Beyond characterizing the crystal structure and phase transitions, a series of studies recently demonstrated that high-power excitation of these phonons can control sample morphology and crystal symmetry group transitions (22-24).

While THz TDS based on asynchronous pump-probe such as ECOPS and ASOPS can achieve spectral acquisition rates as high as GHz. Here we show by using a molecular crystal which has a sufficiently high activation energy that THz TDS can provide a stand-off method to follow rapid transitions, revealing kinetics and underlying bond breaking during the transition.

Walther et al. first explored the temperature dependent anharmonicity of molecular crystal phonons using THz measurements of microcrystalline fructose, sucrose, and glucose powders. While the data was limited for glucose and fructose, sucrose phonons were found to continuously red-shift for temperatures above 270K. These did not examine the behavior at the melting point and did not identify the differences in behavior for different symmetries (25).

Fructose is often used as a calibration standard (26) or for new instrumentation validation (27-29). Recently we introduced single crystal sucrose as a THz linear dichroism standard (30, 31). Similarly, we find strong linear dichroism in fructose which can be used to fill in additional frequencies in THz polarimetry.

The temperature dependence of lattice vibrations for a broad range of materials are readily characterized with THz TDS (4, 32). A number of groups have used THz TDS to characterize solid state phase transitions, however these studies did not examine phonon amplitude changes or anharmonicity to characterize the transitions (2, 5-7, 32, 33). It is likely this is due to the sample preparation either starting from an amorphous melt or from pressed pellets of polycrystalline powders. In either case the disorder and/or grain size of the polycrystalline phases leads to scattering and other effects which result in substantial linewidths obscuring peak shifts.

Nevertheless, the work by Sibik et al. provides an interesting comparison to the data presented here. They measured temperature-dependent THz absorbance of amorphous melts of paracetamol (MW = 151.163 g/mol) showing both crystallization and then transformation to different polymorphs as the temperature increased (7). While the spectra of the three polymorphs are distinct, the features are relatively broad and line fitting was not used. Instead, the amorphous phase versus crystalline phase were estimated from the net absorbance at a specific temperature and frequency. The Avrami-Erofeev fit was then done using the temperature axis as an effective time, given a constant heating rate, resulting in an effective amorphous to crystallization rate of k = 0.056 min$^{-1}$ with heating rate of 0.4K/min. It is difficult to compare this value to the one we extract as they examined the amorphous melt to crystalline transition, rather than the melting. In addition, our fit is to an isothermal change whereas theirs is for a continuous change in the temperature. Nevertheless, the fact that their value is within a factor of 2 for a molecular crystal with similar molecular weight molecule does perhaps help to set the general values one might expect.

In these measurements we used "steady state" THz TDS to monitor the melting transition, where the spectra are collected over ~ 1 min each. Even with this slow acquisition time, we can follow the loss in structure. The very long melting time is well known for many molecular crystals and has enabled the determination of temperature dependence of the melting rate where the transition can be followed at soaking temperatures in excess of the nominal melting temperatures. Lower molecular weight crystals such as isobutene and methyl benzene, required fast scanning calorimetry (FSC) with a 1 μs time resolution (34), but with larger molecular weight glucose and fructose, the temperature dependent melting could be measured with standard differential scanning calorimetry (DSC) (8). The low molecular weight molecule crystal FSC measurements by Cubeta et al. found surprisingly large activation energies ~ 100-200 kJ/mol, which were suggestive of highly cooperative processes similar to glass softening (34). That same year, Liavitskaya et al. used standard DSC to examine the kinetics of larger molecular weight molecular crystals of glucose and fructose. As done by Cubeta et al, Liavitskaya et al. found that the nature of the melting transition was best described by a heterogenous growth of the phase rather than homogeneous formation of nuclei. The activation energies were again determined to be large and were analyzed using Mott's estimation of number of molecules involved collectively in the transition from the ratio of the activation energy to the heat of enthalpy. In the case of fructose, they found at least 5 molecules collectively participate in melting, with the synchronous breaking of as many as 17 hydrogen bonds.

The THz measurements of single crystals allow us to both assign the resonances to specific atomic displacements by the anisotropy, but also provide sufficiently sharp resonances so that line fitting can be used to provide insight to the structural changes and extract kinetic parameters. Even while DFT calculations become more accurate, resonant frequencies are often substantially different than the measurements, and anisotropy measurements of single crystals can remove uncertainty in the assignment of the resonances. The excellent agreement with the Avrami-Erofeev model in Fig. 7 suggests the overall approach can provide reliable kinetics data. Ideally there would be several different soaking temperatures used for the THz melting measurements to achieve a fully independent determination of the activation energy. However, the heating rate was too slow to reach higher temperatures before substantial melting. This can be addressed by a change in the design of the sample holder. The slow melting rate of 0.023 min$^{-1}$ at 376K is consistent with the Liavitskaya results, including the large activation energy. Examining the displacement diagrams in Fig. 3 we see the 1.3 THz vibration is has large oxygen displacements which impact the intermolecular hydrogen bonds that would need to break as the system loses long range order. This may be why an initial reconfiguration along the *c*-axis occurs before the melting. These vibrations are both intra and intermolecular in nature (35).

## CONCLUSIONS

Here we show the use of THz spectroscopic characterization of molecular crystal melting. We measure the anisotropic absorbance for fructose and identify the phonons based on their symmetry axis. We find the 1.3 THz and 2.13 THz phonons are aligned along the *c*-axis, the 1.70 THz phonon is aligned along the *b*-axis and the 1.76 THz phonon is aligned along the *a*-axis. As THz polarimetry techniques develop there is a demand for linear dichroism standards. Fructose anisotropy can provide additional characterization of at frequencies complimentary to sucrose resonances. Our melting measurements show that all phonons red shift as the temperature increases, however for the lowest frequency phonon the *c*-axis, the frequencies blue shift, suggesting a structural change at the melting temperature. The melting follows the 3D growth with a rate constant of 0.023 min$^{-1}$ at 376K. The results are consistent with previous suggestions of melting transition involving the collective breaking of bonds. Beyond these results specific for fructose and molecular crystals, the overall method of monitoring the transition using THz TDS offers the possibility of monitoring even those transition that occur with much shorter time constants without needing ultrafast DSC. As single-shot THz TDS has already been established, the time resolution of THz TDS phase transition measurements can readily reach the ns time scale with current GHz pump lasers.

## ACKNOWLEDGMENTS

AD prepared samples, built equipment, performed measurements, analysis and wrote manuscript, YD and DKG performed THz TDS measurements. AGM conceived measurements, prepared samples and wrote manuscript. FV supervised by EE performed measurements, ES supervised by JB performed X-ray indexing and SD supervised by TK performed DFT calculations and renderings. All authors contributed to the preparation of manuscript. This work supported by NSF grants DBI 1556359 and MCB 1616529, Los Alamos National Laboratory and NSF grant DMR 2003932.

Theory: Polymorphs I and III of Diflunisal. *J. Phys. Chem. B* **2016,** *120* (8), 1698-1710.